# MCNP Simulation to Hard X-Ray Emission of KSU Dense Plasma Focus Machine.

**Amgad E. Mohamed**

*Mechanical and Nuclear Engineering Dept., Kansas State University, Manhattan, KS 66502, USA.*

**Abstract**: the MCNP program used to simulate the hard x-ray emission from KSU dense plasma focus device, an electron beam spectrum of maximum energy 100 keV was used to hit anode target. The bremsstrahlung radiation was measured using the F2 tally functions on the chamber walls and on a virtual sphere surrounding the machine, the radiation spectrum was recorded for various anode materials like tungsten, stainless steel and molybdenum. It was found that tungsten gives the best and the most intense radiation for the same electron beam. An aluminum filter of thickness 2mm and 4mm was used to cutoff the lower energy band from the x-ray spectrum. It was found that the filters achieved the mission and there is no distinct difference in between.

**INTRODUCTION:**

plasma focus devices considered one of the best devices to produces intense hard, soft X-ray and neutron bursts which can be used in various medical and industrial applications [1-2], The high energy X-rays come from a small region of the anode surface on the axis due to the collision of an electron beam figure (1), the electron beam energy measured in various experimental works was found to be a spectrum with 50keV most probable energy, with high energy tail around 100 keV, using neon gas as working gas for the device. In this work we simulate the x-ray emission from the Kansas State University Dense Plasma Focus (KSU-DPF) machine, this machine is 10kJ, with working energy at 1.6kJ at charging voltage of 17kV. The anode length of the machine is 10cm with 1.5 cm diameter [3-5].

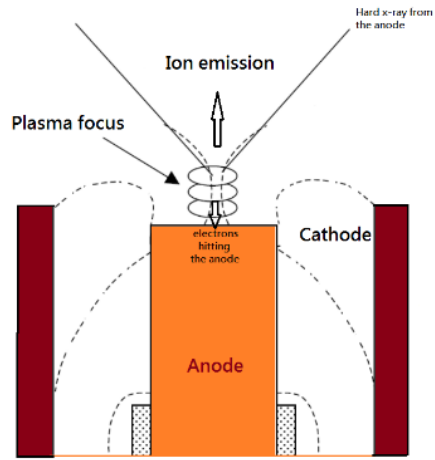

**Figure 1: X-ray emission from the dense plasma focus.**

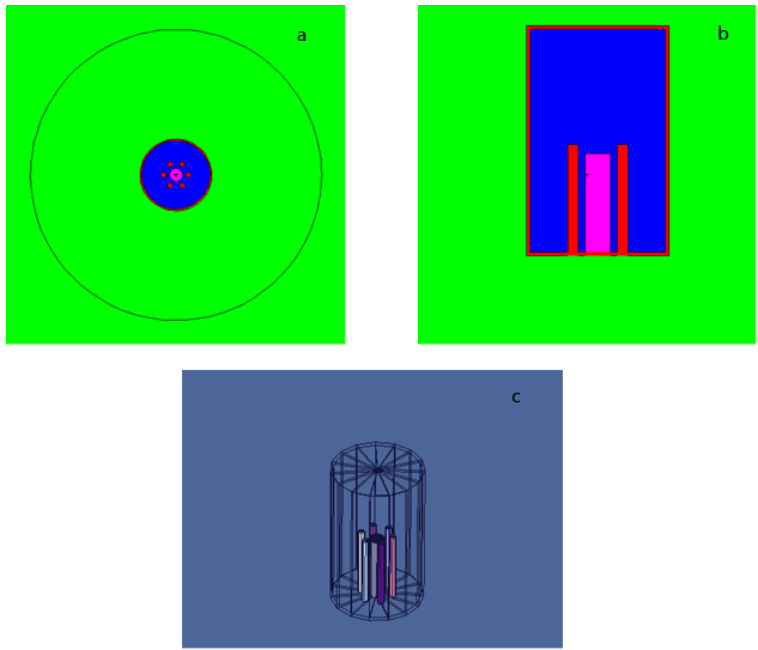

**Fig 2: (a) x-y plane for the plasma focus chamber, (b) x-z plane for the plasma focus chamber, (c) 3D projection for the experiment**

**SIMULATION PROCEDURE**

    The MCNP was used [6] to build the plasma focus device with a cylindrical chamber of stainless steel contains one anode in the middle and 6 cathodes around figure 2.

A source of electron beam was investigated with energy 100 keV above the anode and directed downward to hit the anode material. A sphere with radius greater than the cylinder height was established around the device to measure the flux of photons figure 3.

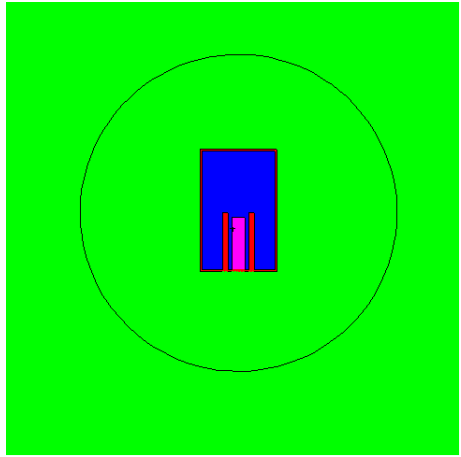

**Figure 3: the sphere used to measure the flux**

The number of histories was set to nps=1e5, and the physics card included the bremsstrahlung radiation.  (Phys:E 0.5 0 0 0 0 1. 1. 1. 1. 1)

**X-RAY SIMULATION RESULTS**

The electron scattering and photon emission was plotted by using particle display function in the visual editor [7] figure 4. The electrons were found to scatter at max 60 degree from the vertical axis.

The photons tracks were noticed to have greater angle ~ 80 degree with the vertical axis.

the F2 tally results indicating the spectrum of the x-ray were plotted and it's clear that the maximum energy in the tail of the spectrum equal to electron beam energy, the noticed pulses in the spectrum are from the k and l electron shells of the anode materials.

the next figure 6 indicates the bremsstrahlung radiation on the inside surface of the chamber from tungsten anode material with density 19.3 g/cm3, the k and l shell energies from the scattering on the stainless steel chamber wall are very clear at at 5.7 and 6.8 keV. The spectrum from stainless steel anode of density 7.8 g/cm3 was plotted in figure 7 , it is found that the number of photons emitted is 5 times lower than the tungsten use. By using a molybdenum anode of density 10.22 g/cm3 the photon number is greater than the stainless steel and lower than the tungsten, figure 8.

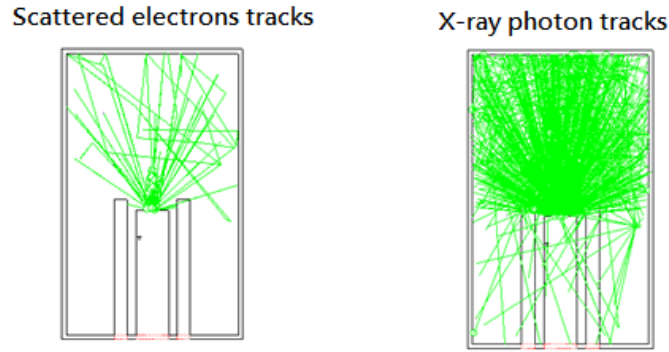

**Figure 4: (Left), The electrons scattring from the anode; (Right), The photons emission track from the anode material.**

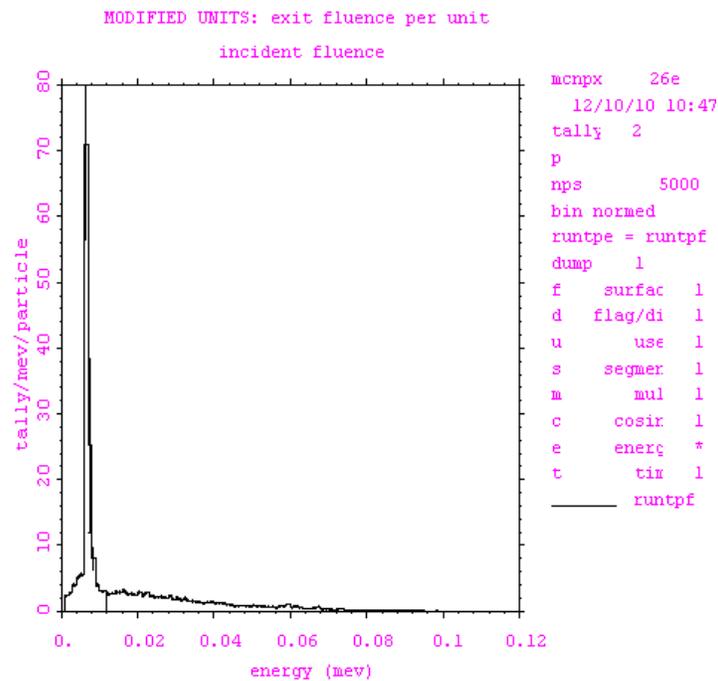

**Figure 6: the bremsstrauhlung spectrum from tungsten anode, the k and l shell energy from the stainless steel chamber at 5.7 and 6.8 keV**

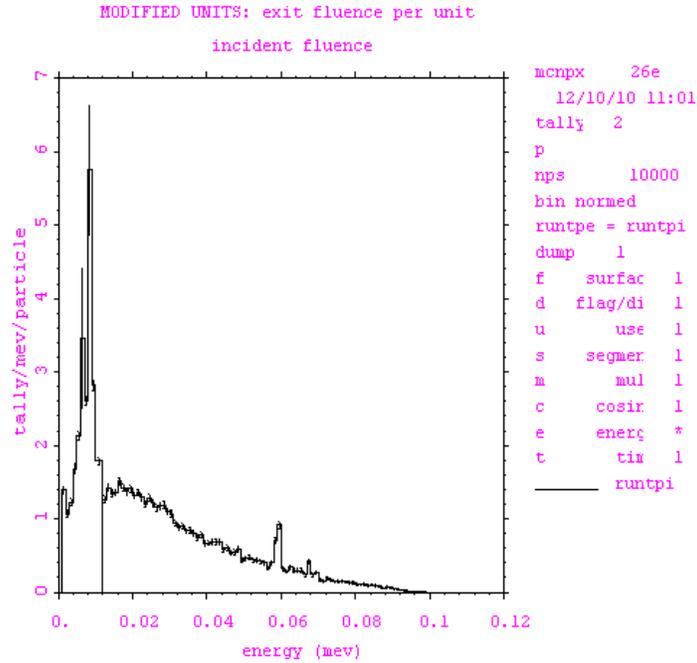

**figure 7: the bremsstrauhlung spectrum from stainless steel anode, the k and l shell energy from the stainless steel at 5.9 and 6.5 keV**

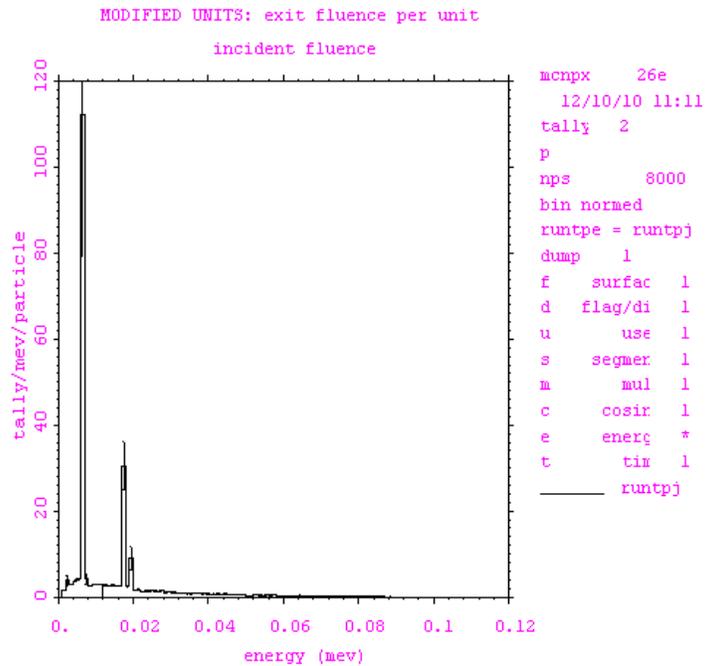

**figure 8: the bremsstrauhlung spectrum from molybdenum anode, k, and l lines at 19.6 and 17.5keV.**

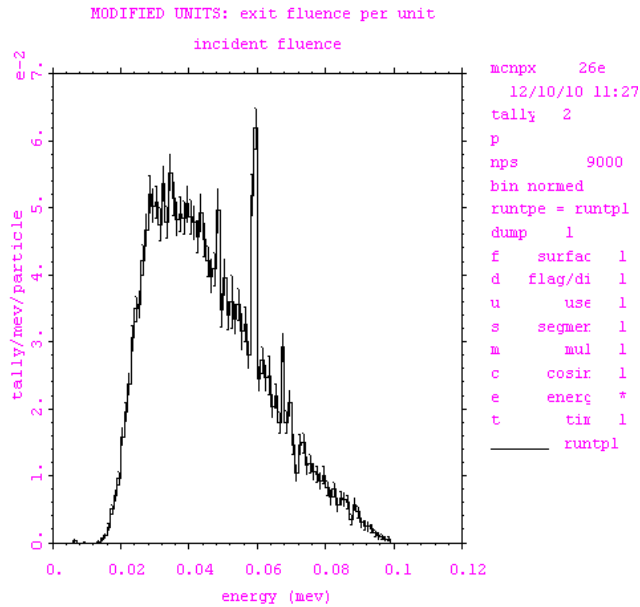

**Figure 9 : the spectrum from tungestun using 2 mm thick aluminum filter, lines at 57 keV and 68 keV.**

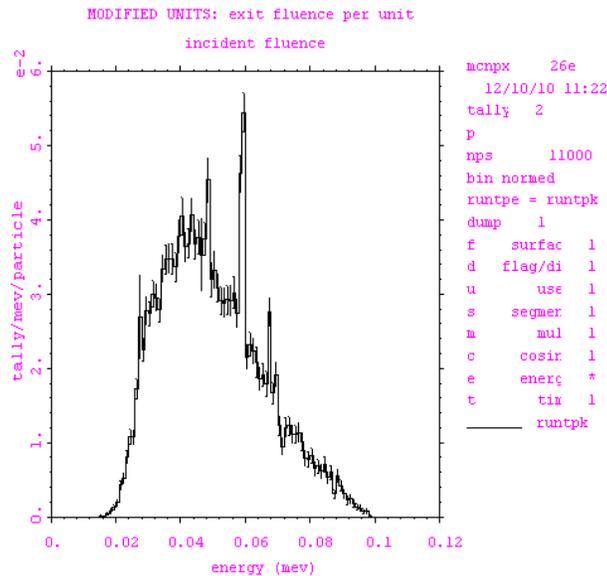

**Figure 10 : the spectrum from tungestun using 4 mm thick aluminum filter.**

By using a filter of aluminum sheet with density 2.7 g/cm3 and thickness 2mm and 4 mm. the spectrum was measured on the outer sphere, found to cut the lower energy part which is in great benefits in medical scan as the lower energy is very harmful to the human body figure 9, 10. The photon number was not greatly affected by changing the thickness from 2 to 4 mm.

The k and l shell energies for the tungsten was clear to be noticed at 57 keV and 68 keV.
The k and l shell from the stainless steel were cutoff.

**CONCLUSION AND FUTURE WORK:**

The MCNP was found to be an excellent method to predict the x-ray emission from the plasma focus machine, with different anode materials. Using filters can control the band of energy for the x-ray needed for the appropriate application.

The best emission intensity was found from the tungsten anode which gives the larger bremsstrahlung radiation. In the future work the MCNP will be used to simulate the neutron and proton interaction in the machine to characterize the whole radiation types from the KSU-DPF machines.